\documentclass{emulateapj}

\begin{document}
\title{A HERO'S DARK HORSE: DISCOVERY OF AN ULTRA-FAINT MILKY WAY SATELLITE IN PEGASUS}
\author{Dongwon Kim} 
\author{Helmut Jerjen} 
\author{Dougal Mackey} 
\author{Gary S. Da Costa} 
\author{Antonino P. Milone}
\affil{Research School of Astronomy and Astrophysics, The Australian National University, Mt Stromlo Observatory, via Cotter Rd, 
Weston, ACT 2611, Australia}

\email{dongwon.kim@anu.edu.au}

\begin{abstract}
We report the discovery of an ultra-faint Milky Way satellite galaxy in the constellation of Pegasus. The concentration of stars was 
detected by applying our overdensity detection algorithm to the SDSS-DR 10 and confirmed with deeper photometry 
from the Dark Energy Camera at the 4-m Blanco telescope. Fitting model isochrones indicates that this object, 
Pegasus\,III, features an old and metal-poor stellar population ([Fe/H]$\sim-2.1$) at a heliocentric distance of $205\pm20$\,kpc. 
The new stellar system has an estimated half-light radius of $r_h=78^{+30}_{-24}$\,pc and a total luminosity of $M_{V}\sim-4.1\pm0.5$ that
places it into the domain of dwarf galaxies on the size--luminosity plane. Pegasus\,III is spatially close to the MW satellite Pisces\,II.
It is possible that the two might be physically associated, similar to the Leo\,IV and Leo\,V pair. Pegasus\,III is also well aligned with the Vast Polar Structure, which suggests a possible physical association.

\end{abstract}

\keywords{Local Group -- Milky Way, satellites: individual: Pegasus III}

\section{Introduction}
Following the Sloan Digital Sky Survey~\citep{York2000}, more recent wide-field imaging surveys such as the Stromlo Milky Way Satellite survey~\citep{SMS}, the Dark Energy Survey~\citep{DES}, the Pan-STARRS 3$\pi$ Survey (K. Chambers et al., in preparation), and the Survey of the Magellanic Stellar History (SMASH;  PI D. Nidever)  have been revealing new Milky Way companions including satellite galaxies~\citep{Koposov2015,Bechtol2015,Laevens2015,Martin2015} and star clusters~\citep{Belokurov2014,Laevens2014,Kim1, Kim2}. The new Milky Way companions, many of which are in the southern sky, share the properties of previously discovered ultra-faint stellar systems, such as low luminosities ($ -8 \lesssim M_{V} \lesssim -1.5 $)~\citep{Martin2008} and low metallicities [Fe/H]$<-2$~\citep{Kirby2008,Norris2010,Simon2011,Koch2014}. 

In this letter we report the detection of the new ultra-faint Milky Way satellite Pegasus\,III (Peg\,III) found in SDSS Data Release 10 and confirmed with deep DECam imaging (Sections 2 \& 3). Peg\,III appears to be located at a heliocentric distance of $\sim205$ kpc and have a half-light radius of $\sim78$ pc (Section 4). In the last section we discuss the possible origin of the new satellite galaxy and conclude with our results.

\section{Discovery}

The SDSS is a photometric and spectroscopic survey in the $ugriz$ photometric bands to a depth of $r\sim22.5$ magnitudes~\citep{York2000}. Data Release 10 (DR10), publicly available on the SDSS-III Web site\footnote{http://www.sdss3.org/dr10/}, covers $14,555 \deg^{2}$ mostly around the north Galactic pole~\citep{Ahn2014}.

 The new object was first flagged by our detection algorithm in the search for stellar overdensities over the existing SDSS catalog as described in ~\cite{Invisibles} and \cite{Kim1}. Briefly, we used isochrone masks based on the PARSEC stellar evolution models~\citep{Parsec} as a photometric filter to enhance the presence of old and metal-poor stellar populations relative to 
the Milky Way foreground stars. We then binned the R.A., DEC. positions of the filtered stars and convolved the density-map with a Gaussian kernel. Based on the density-map, we calculate the signal to noise ratios (S/Ns) of potential overdensities and measure their significance by comparing their S/Ns to those of random clusters in the residual background. Moving the isochrone masks over a range of distance moduli $(m-M)$ between 16 and 22 magnitudes, this process is repeated with different scales of bins and Gaussian kernels.

With this algorithm, we recovered all of the previously known MW companions in the SDSS coverage and found a few more promising candidates, one of which was reported in~\cite{Kim1}. The new object was detected with a significance of $\sim7\sigma$ in the constellation of Pegasus.

\section{Follow-up Observations and Data Reduction}
Deeper follow-up observations of the Peg\,III field were conducted on during the night of 17th July 2014 using the Dark Energy Camera (DECam) at the 4-m Blanco Telescope located at Cerro Tololo Inter-American Observatory (CTIO) in Chile. DECam imager is equipped with a focal plane array containing sixty-two 2k $\times$ 4k CCD detectors with a wide field of view (3.0 square degrees) and a pixel scale of $0\farcs27$ (unbinned). Under  photometric conditions, we obtained 840\,s exposures in $g$ and 1050\,s  in $r$ band, divided over dithered single exposures of 210\,s. The average seeing during the observing was $1\farcs3$ in the $g$ and $1\farcs1$ in the $r$ band. The single exposure images were fully reduced and stacked through the DECam community pipeline \citep{DECamCP2014}. We carried out weight-map combination, source extraction and PSF photometry with the use of WeightWatcher \citep{WeightWatcher} and SExtractor/PSFEx \citep{SExtractor,PSFEx}. For star/galaxy separation, we applied the threshold $\mathtt{\left| SPREAD\_MODEL\right|<0.003+SPREADERR\_MODEL}$ as described in \cite{Koposov2015}. We then positionally matched the star-like objects with SDSS stars with a maximum radius of $1\farcs0$ using the STILTS software \citep{STILTS} . We used this catalog for the photometric calibration in the magnitude range $17.0<r_{0}<21.0$ mag, between the saturation limit of the DECam and the $5\sigma$ limit of the SDSS. Finally, the magnitudes of the calibrated point sources were dereddened by means of the \cite{Schlegel1998} extinction maps and the extinction correction coefficients of~\cite{Schlafly2011}.

\begin{figure}
\begin{centering}
\includegraphics[scale=0.6]{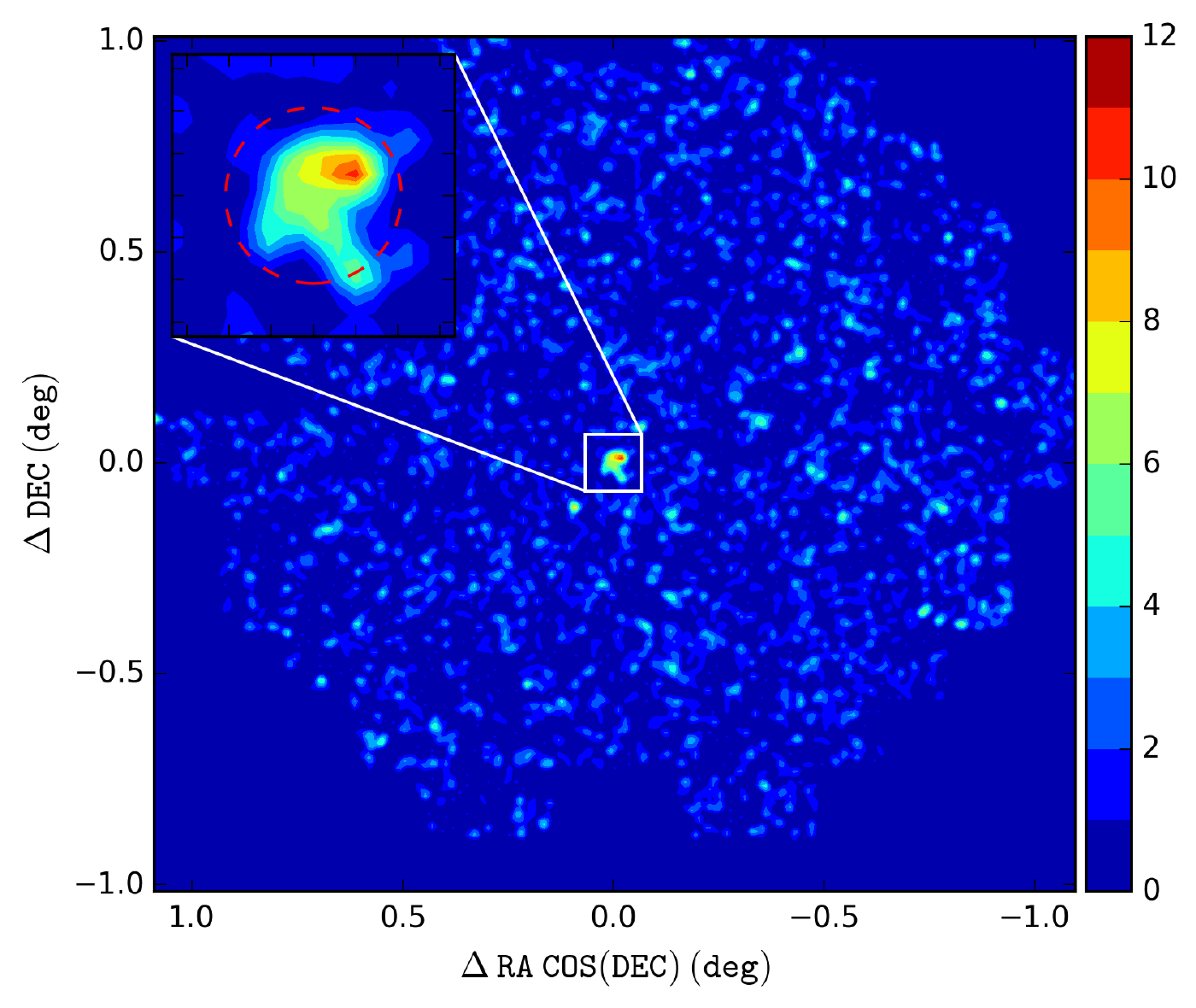}
\includegraphics[scale=0.6]{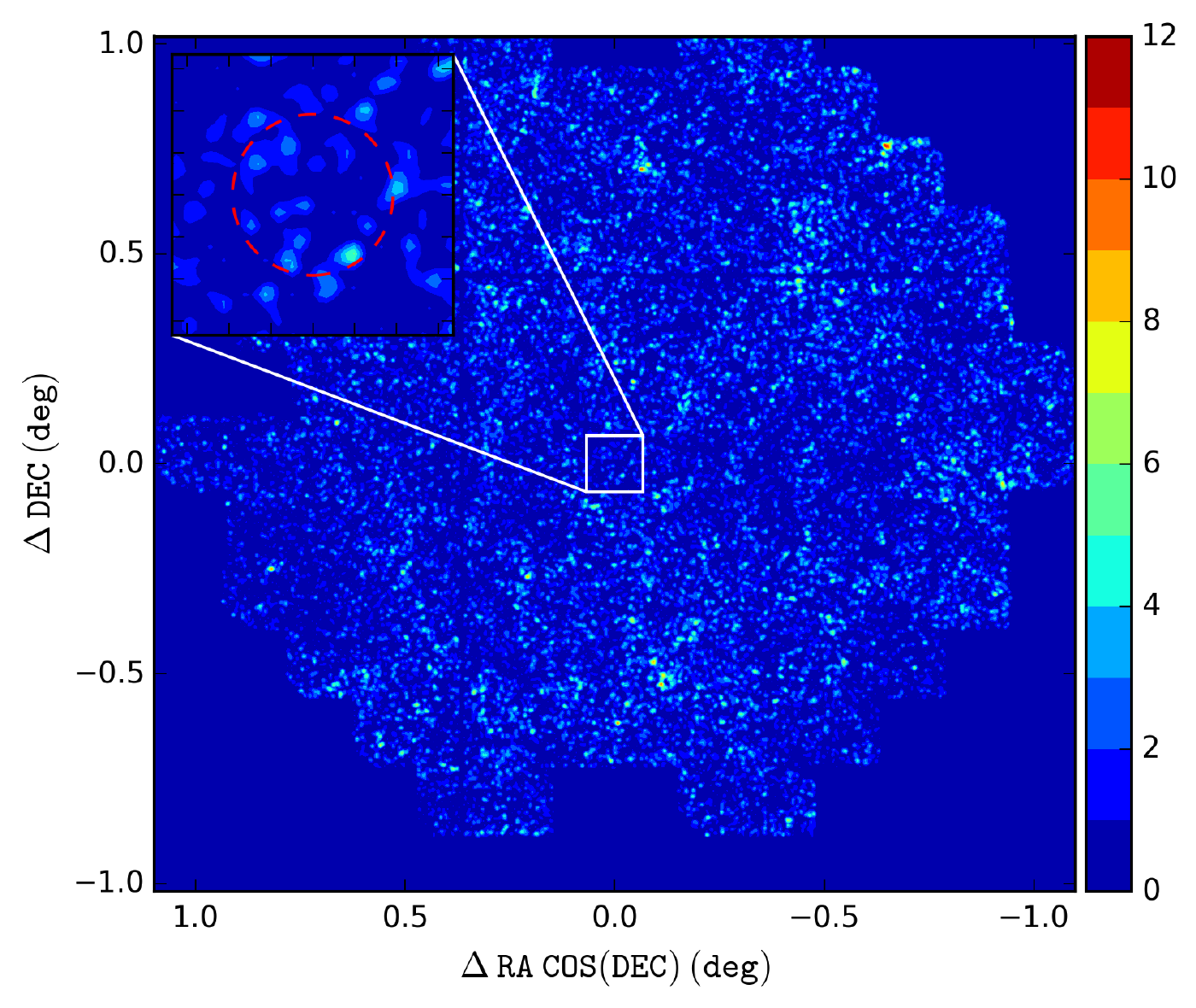}
\par\end{centering}
\caption{Upper panel: Density contours of candidate stars in the field of view of DECam that pass the photometric filter with a PARSEC isochrone of 13.5 Gyr and [Fe/H]$=-2.1$~\citep{Parsec} at the distance modulus $(m-M)=21.56$ magnitudes. The contours show the levels of star density in units of the standard deviation above the background. The dashed circle marks a radius of $2\farcm5$ (see Section\,4 for details). Lower panel: Same as the upper panel but for all objects classified as galaxies. At the central region, there is no obvious overdensity coincident with that in the upper panel.}
\label{fig:Contour}
\end{figure}

The upper panel of Figure~\ref{fig:Contour} shows the contour map of star density centred on Peg\,III in the field of view of DECam after the photometric filtering process using a PARSEC isochrone of 13.5 Gyr and [Fe/H]$=-2.1$~\citep{Parsec} at the distance modulus $(m-M)=21.56$ magnitudes. High level density contours ($>4\sigma$) clearly define Peg\,III  in the central region of the image and no other comparable overdensity in terms of S/N. Our algorithm recovers the overdensity with a significance of $\sim10\sigma$ in the DECam data. The irregular shapes of the outer isophotes are likely due to the fact that we sample only the brightest red-giant branch (RGB) and horizontal branch (HB) stars in the system or that we see the signature of tidal disturbance. In the lower panel we present the same kind of contour map as shown in the upper panel but for all galaxy-like objects classified by the threshold $\mathtt{SPREAD\_MODEL>0.003}$. There is no overdensity coincident with Peg\,III, ruling out the possibility of a background galaxy cluster.

\section{Candidate Properties}

\begin{figure*}[t!]
\includegraphics[scale=0.7]{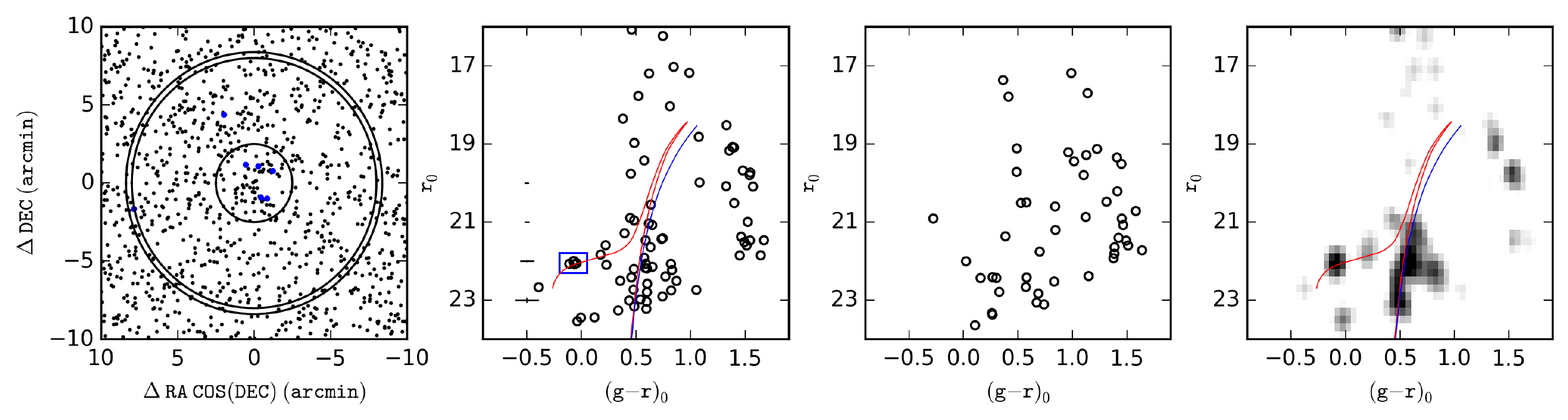}
\caption{DECam view of Peg\,III. Left panel: Distribution of all objects classified as stars in a $20' \times20'$ field that shares the central coordinates with Figure~\ref{fig:Contour}. The circles mark a radius of $2\farcm5$ (equal to the dashed circle in Figure~\ref{fig:Contour}), $8\farcm0$ and $8\farcm3$ respectively. The blue dots mark the seven BHB stars that fall into the color-magnitude range $-0.20<(g-r)_{0}<0.05$ and $21.80<r_{0}<22.30$, illustrated as a blue box in the next panel. Middle left panel: CMD of stars lying within the inner circle, dominated by the candidate stars of the dwarf galaxy. Middle right panel: Comparison CMD of stars lying in the annulus defined by the two outer circles, dominated by foreground stars. Right panel: Field-subtracted Hess diagram, the inner CMD minus the comparison CMD, showing an excess of stars at the locations of the BHB and RGB of Peg\,III. The best-fitting PARSEC (red) isochrone of age 13.5 Gyr and [Fe/H]=-2.1 and Dartmouth~\citep{Dartmouth} (blue) isochrone of age 14.2 Gyr and [Fe/H]$=-2.3$ are overplotted at a distance of 205 kpc.\label{fig:HessDiagram}}
\end{figure*}

The left panel of Figure~\ref{fig:HessDiagram} shows the RA-DEC distribution of all stellar objects identified by SourceExtractor in the vicinity of Peg\,III. The middle left panel of Figure~\ref{fig:HessDiagram} shows the extinction-corrected CMD of stars within $2\farcm5$ (equal to the dashed circle in Figure~\ref{fig:Contour}) of the nominal centre of Peg\,III, and the middle right panel that of the foreground stars in the same manner as in \citet{Belokurov2010} from annulus defined by the outer radii of $8\farcm0$ and $8\farcm3$ covering the same area as the inner circle around Peg\,III. Finally, the right panel shows a field-subtracted Hess diagram, built on the CMDs in the middle. We note that Peg\,III shares a well defined RGB and blue-HB (BHB) with other distant MW satellite galaxies, namely, Leo\,V and Pisces\,II~\citep{Belokurov2008,Belokurov2010}. The five BHB candidate stars clustering at $r_{0}\sim22$\,mag and $(g-r)_{0}\sim-0.1$\,mag are highlighted as blue dots in the inner circle in the left panel of Figure~\ref{fig:HessDiagram}. Fitting the horizontal branch of the PARSEC isochrone model yields an average heliocentric distance of $\sim205$\,kpc. 

\begin{figure}[t!]
\begin{centering}
\includegraphics[scale=0.85]{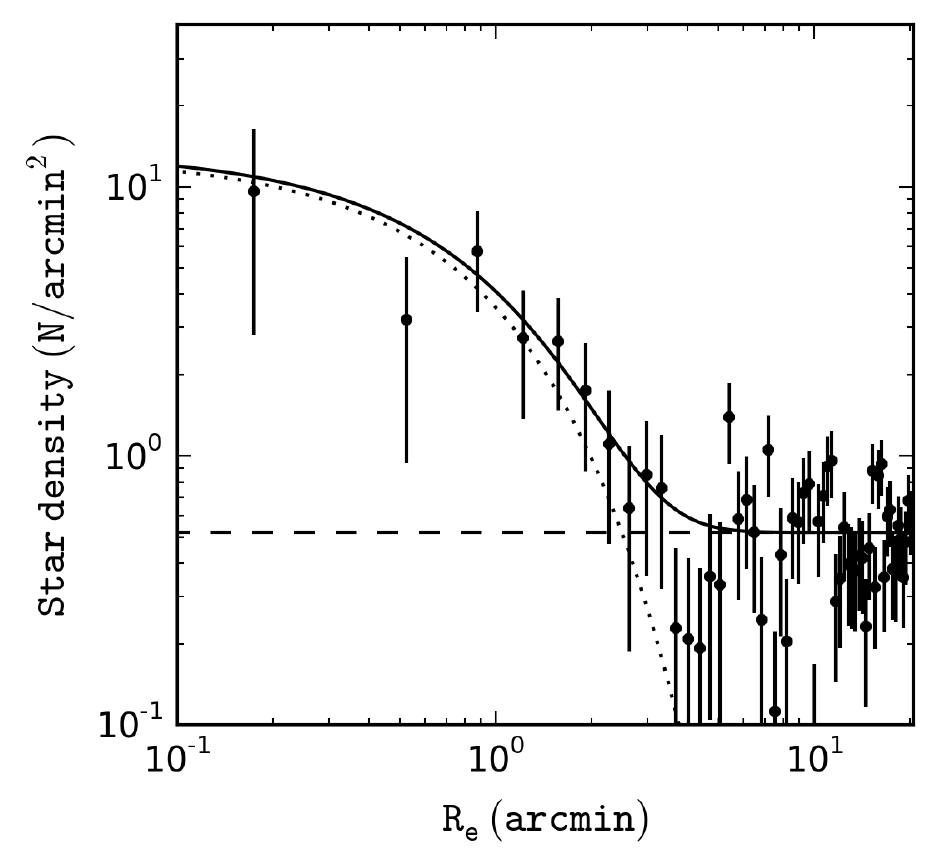}
\end{centering}
\protect\caption{Radial stellar density profile of Peg\,III. $R_{e}$ is the elliptical radius. Overplotted are the best exponential model based on the parameters in Table~\ref{tab:Parameters} (dotted), the contribution of foreground stars (dashed) and the combined fit (solid). The error bars were derived from Poisson statistics.\label{fig:RadialProfile}}
\end{figure}

To derive the central position $\alpha$ and $\delta$, ellipticity $\epsilon$,  position angle $\theta$ and half-light radius $r_{h}$, we employed a maximum-likelihood algorithm similar to the procedure described in~\cite{Martin2008} with the stars that passed the photometric filter. We obtained a half-light radius of $1.3^{+0.5}_{-0.4}$\,arcmin, or  $78^{+30}_{-24}$\,pc adopting a heliocentric distance of 205\,kpc. The radial density profile with the best-fitting exponential model is presented in Figure~\ref{fig:RadialProfile}.
 
We estimate the total luminosity of Peg\,III in analogy to \cite{Walsh2008} as follows. First, we calculate the total number of Peg\,III stars $N_{*}$ within the photometric limit by integrating the best-fitting exponential profile shown in Figure~\ref{fig:RadialProfile}. We use the ratio of the total number of stars $N_{*}$ to the probability density of a normalised theoretical luminosity function (LF) in the same magnitude range. Using this ratio, we scale the theoretical LF to the star number density as a function of $r$ magnitude. We then integrate the scaled LF taking into account the missing flux beyond the lower limit of our photometry and obtain a total luminosity $M_{r}=-4.23$\,mag based on the initial mass function by~\cite{Kroupa2001} and PARSEC isochrone for a 13.5 Gyr old stellar population with [Fe/H]$=-2.1$. From the $V-r=0.17$\,mag luminosity weighted mean color for the model isochrone we derive the corresponding $V$-band luminosity $M_{V}=-4.1$\,mag. Since this method relies on total star counts instead of individual flux, the inclusion or exclusion of a single RGB in the system carries large uncertainties up to $\sim0.5$ mag. Hence, a realistic estimate of the total luminosity of Peg\,III is $M_{V}=-4.1\pm0.5$. All derived parameters are summarised in Table~\ref{tab:Parameters}. 

\begin{deluxetable}{lrl}
\tablewidth{0pt}
\tablecaption{Properties of Peg\,III}
\tablehead{
\colhead{Parameter} & 
\colhead{Value} &
\colhead{Unit}}
\startdata
$\alpha_{J2000}$ & 22 24 22.6$\pm$1.0 & h m s \\
$\delta_{J2000}$ & +05 25 12$\pm$14 & $^\circ$ $\arcmin$ $\arcsec$ \\
$l$ & 69.852 & deg\\
$b$ & $-$41.813 & deg\\
$(m-M)$ & $21.56\pm0.20$ & mag \\
$d_\odot$ & $205\pm20$ & kpc \\
$r_{h}$ & $78^{+30}_{-24}$ \tablenotemark{a} & pc \\
$\epsilon$ & $0.46^{+0.18}_{-0.27}$ & \\
$\theta$ & $133\pm17$ & deg \\
$M_{tot,V}$ & $-4.1\pm0.5$ & mag \\
\enddata
\tablenotetext{a}{ Adopting a distance of 205\,kpc}
\label{tab:Parameters}
\end{deluxetable}

\begin{figure}[t!]
\begin{centering}
\includegraphics[scale=0.45]{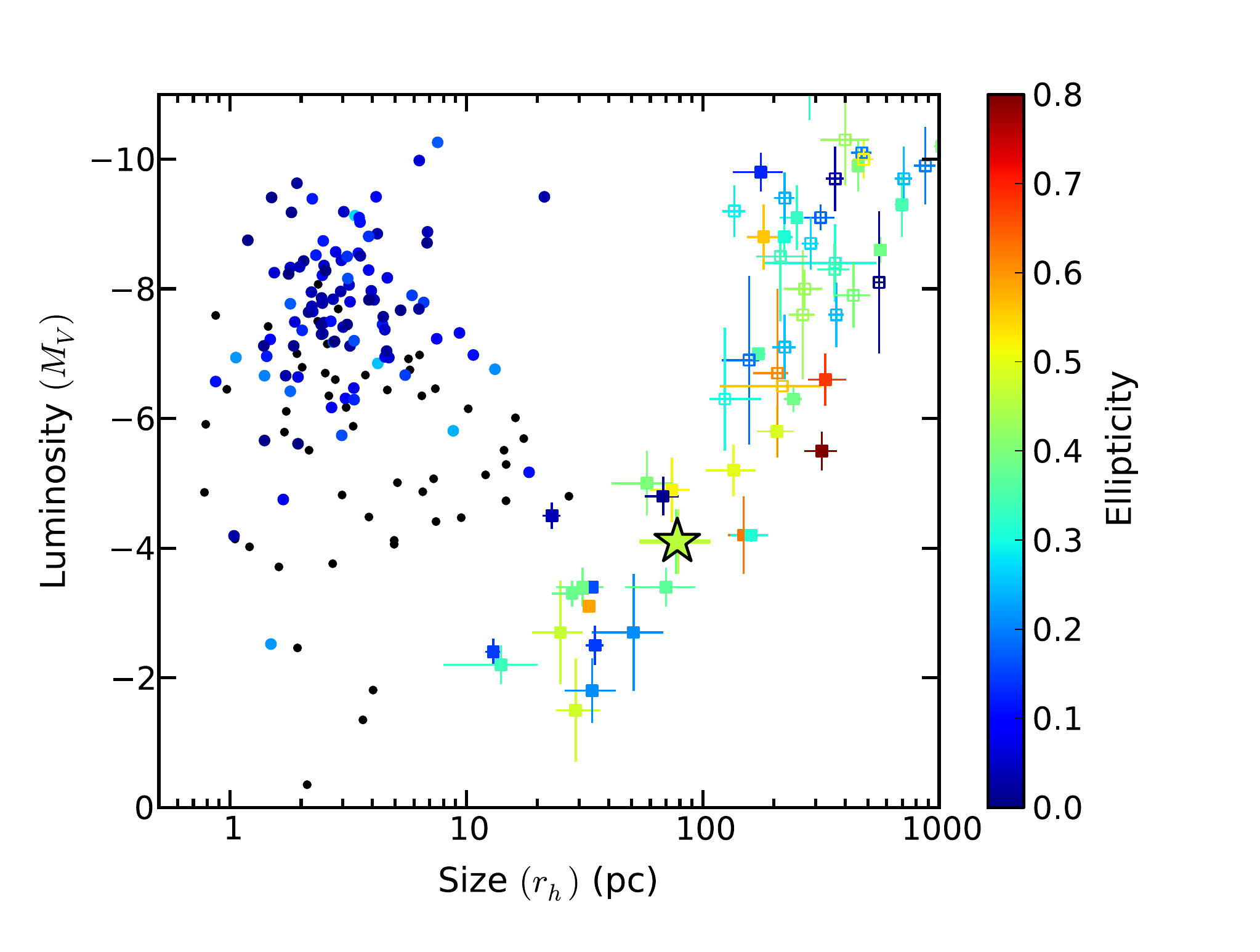}
\end{centering}
\protect\caption{The position of Peg III on the size-luminosity plane, marked with a star outlined in black.  Also shown are all Milky Way globular clusters (filled circles), and the presently-known dwarf spheroidal satellites of the Milky Way (filled squares) and M31 (open squares).  All points are colour-coded by ellipticity; those globular clusters lacking an ellipticity measurement are marked in black. Peg III clearly occupies the region inhabited by ultra-faint dwarf satellites of the Milky Way, and it has a comparable ellipticity to many of these objects. We note that Peg III directly falls on top of Coma Berenices, which is therefore invisible on this plot. Measurements for the globular clusters were taken from~\citet[and 2010 edition]{Harris1996}; those for the Milky Way dwarfs from \cite{McConnachie2012}, \cite{Belokurov2014},\cite{Laevens2014, Laevens2015}, \cite{Kim2}, \cite{Koposov2015}, \cite{Bechtol2015}, and \cite{Martin2015}; and those for the M31 dwarfs from \cite{McConnachie2012}, and \cite{Martin2013a,Martin2013b} In the case where more than one independent set of luminosity and/or structural measurements exists for an object, we adopt their weighted mean.\label{fig:size-lum-ellip}}
\end{figure}

\section{Discussion and Conclusion}
We report the discovery of a new ultra-faint Milky Way satellite, Peg\,III. The satellite hosts a typical old, metal-poor stellar 
population as it is observed in many other Milky Way satellite galaxies. Its large half-light radius ($\sim 78$\,pc) and 
luminosity ($-4.1\pm0.5$) puts Peg\,III among other systems classified as ultra-faint dwarf galaxies in the size-luminosity 
parameter space as shown in Figure~\ref{fig:size-lum-ellip}. In particular, these physical properties of Peg\,III are very similar to those of previously known remote MW satellites such as Leo\,IV \citep[$d_\odot=154\pm5$\,kpc, $r_{h}=206^{+36}_{-31}$\,pc;][]{Moretti2009}, Leo\,V \citep[$d_\odot=196\pm15$\,kpc, $r_{h}=65\pm30$\,pc;][]{Sand2012} and Pisces\,II~\citep[$d_\odot=183\pm15$\,kpc, $r_{h}=58\pm10$\,pc;][]{Sand2012}. We note that we could have overestimated the half-light radius of Peg\,III due to the sampling of stars limited to RGB/HB and field-contamination. Deeper photometry down to the main-sequence stars tends to give smaller half-light radii as shown in previous studies~\citep[e.g. Leo\, V in ][]{deJong2010,Sand2012}. However, even if the true size is only half of the current estimate, Peg\,III would be still found in the region where ultra-faint dwarf galaxies populate in the size-luminosity parameter space.

It is interesting to note that Peg\,III and Pisces\,II are separated only by 8.5\,deg on the sky and have fairly similar distances ($205\pm20$\,kpc 
and $183\pm15$\,kpc). The spatial separation of the two satellite galaxies is $\sim 30$\,kpc. A similar situation has already been encountered 
with the Leo\,IV and Leo\,V pair~\citep{deJong2010}. This suggests that Peg\,III and Pisces\,II could be associated with each other, although a velocity measurement will be required to confirm or reject this idea. As for the Leo\,IV--Leo\,V pair they might be related to a single disrupting or disrupted progenitor. It might be a pure coincidence but the Peg\,III--Pisces\,II and Leo\,IV--Leo\,V pairs are almost diametrically opposite (in fact 162\,deg) in the sky, and have almost the same angular separation of $\sim 90$\,deg from the barycenter of the Magellanic Clouds, ie. 88\,deg and 103\,deg, respectively.

With an angular distance of $\sim 13$\,deg, Peg\,III lies also close the vast polar structure (VPOS), a planar arrangement defined by the 
27 previously known Milky Way satellite galaxies, including the Magellanic Clouds, perpendicular to the MW disk \citep{Kroupa2005,Metz2007,Metz2009,Kroupa2010,Pawlowski2012}.
The majority of recently discovered Milky Way satellite candidates in the southern hemisphere \citep{Koposov2015,Bechtol2015} are also well 
aligned with the VPOS (Pawlowski et al. 2015, in preparation). The origin of that plane is still a matter of debate. It could be the result of a major 
galaxy collision that left debris in form of tidal dwarfs and star clusters along the orbit~\citep{Pawlowski2013}. Peg\,III might be part of that debris.

\acknowledgements{The authors like to thank Tammy Roderick, Kathy Vivas and David James for their assistance during the DECam observing run. We also thank the referee Vasily Belokurov for the helpful comments and 
suggestions, which contributed to improving the quality of the publication. We acknowledge the support of the Australian Research Council through Discovery project DP150100862 and Discovery Early Career Researcher Award DE150101816, and financial support from the Go8/Germany Joint Research Co-operation Scheme. Funding for SDSS-III has been provided by the Alfred P. Sloan Foundation, the Participating Institutions, the National Science Foundation, and the U.S. Department of Energy Office of Science. The SDSS-III web site is http://www.sdss3.org/.

SDSS-III is managed by the Astrophysical Research Consortium for the Participating Institutions of the SDSS-III Collaboration including the University of Arizona, the Brazilian Participation Group, Brookhaven National Laboratory, Carnegie Mellon University, University of Florida, the French Participation Group, the German Participation Group, Harvard University, the Instituto de Astrofisica de Canarias, the Michigan State/Notre Dame/JINA Participation Group, Johns Hopkins University, Lawrence Berkeley National Laboratory, Max Planck Institute for Astrophysics, Max Planck Institute for Extraterrestrial Physics, New Mexico State University, New York University, Ohio State University, Pennsylvania State University, University of Portsmouth, Princeton University, the Spanish Participation Group, University of Tokyo, University of Utah, Vanderbilt University, University of Virginia, University of Washington, and Yale University.

This project used data obtained with the Dark Energy Camera (DECam), which was constructed by the Dark Energy Survey (DES) collaborating institutions: Argonne National Lab, University of California Santa Cruz, University of Cambridge, Centro de Investigaciones Energeticas, Medioambientales y Tecnologicas-Madrid, University of Chicago, University College London, DES-Brazil consortium, University of Edinburgh, ETH-Zurich, Fermi National Accelerator Laboratory, University of Illinois at Urbana-Champaign, Institut de Ciencies de l'Espai, Institut de Fisica d'Altes Energies, Lawrence Berkeley National Lab, Ludwig-Maximilians Universitat, University of Michigan, National Optical Astronomy Observatory, University of Nottingham, Ohio State University, University of Pennsylvania, University of Portsmouth, SLAC National Lab, Stanford University, University of Sussex, and Texas A$\&$M University. Funding for DES, including DECam, has been provided by the U.S. Department of Energy, National Science Foundation, Ministry of Education and Science (Spain), Science and Technology Facilities Council (UK), Higher Education Funding Council (England), National Center for Supercomputing Applications, Kavli Institute for Cosmological Physics, Financiadora de Estudos e Projetos, Funda\c{c}\~ao Carlos Chagas Filho de Amparo a Pesquisa, Conselho Nacional de Desenvolvimento Cient\'i­fico e Tecnol\'ogico and the Ministério da Ci\^encia e Tecnologia (Brazil), the German Research Foundation-sponsored cluster of excellence "Origin and Structure of the Universe" and the DES collaborating institutions.}


\end{document}